\documentclass[preprint,superscriptaddress,preprintnumbers,amsmath,amssymb,prd]{revtex4}
\usepackage{graphicx}

\begin{document}

\title{Fluctuation-induced free energy of thin peptide films}

\author{
M. A. Baranov}
\affiliation{Institute of Physics, Nanotechnology and
Telecommunications, Peter the Great Saint Petersburg
Polytechnic University, Saint Petersburg, 195251, Russia}

\author{
G.~L.~Klimchitskaya}
\affiliation{Central Astronomical Observatory at Pulkovo of the
Russian Academy of Sciences, Saint Petersburg,
196140, Russia}
\affiliation{Institute of Physics, Nanotechnology and
Telecommunications, Peter the Great Saint Petersburg
Polytechnic University, Saint Petersburg, 195251, Russia}

\author{
V.~M.~Mostepanenko}
\affiliation{Central Astronomical Observatory at Pulkovo of the
Russian Academy of Sciences, Saint Petersburg,
196140, Russia}
\affiliation{Institute of Physics, Nanotechnology and
Telecommunications, Peter the Great Saint Petersburg
Polytechnic University, Saint Petersburg, 195251, Russia}
\affiliation{Kazan Federal University, Kazan, 420008, Russia}

\author{
E. N. Velichko}
\affiliation{Institute of Physics, Nanotechnology and
Telecommunications, Peter the Great Saint Petersburg
Polytechnic University, Saint Petersburg, 195251, Russia}

\begin{abstract}
We apply the Lifshitz theory of dispersion forces to find a contribution to the free energy of
peptide films which is caused by the zero-point and thermal fluctuations of the electromagnetic
field. For this purpose, using available information about the imaginary parts of dielectric
permittivity of peptides, the analytic representation for permittivity of typical peptide along
the imaginary frequency axis is devised. Numerical computations of the fluctuation-induced free
energy are performed at room temperature for the freestanding peptide films, containing different
fractions of water, and for similar films deposited on dielectric (SiO$_2$) and (Au) substrates. It
is shown that the free energy of a freestanding peptide film is negative and, thus, contributes to its
stability. The magnitude of the free energy increases with increasing fraction of water and decreases
with increasing thickness of a film. For peptide films deposited on a dielectric substrate the free
energy is nonmonotonous. It is negative for thicker than $100~$nm films, reaches the maximum
value at some film thickness, but vanishes and changes its sign for thinner than $100~$nm films. The
fluctuation-induced free energy of peptide films deposited on metallic substrate is found to be positive
which makes films less stable. In all three cases, simple analytic expressions for the free energy
of sufficiently thick films are found. The obtained results may be useful to attain film stability in
the next generation of organic microdevices with further shrinked dimensions.
\end{abstract}

\maketitle

\section{Introduction}
In the last few years organic electronics has assumed great importance because of its
value in many applications~\cite{b1}. For thin protein and peptide films and coatings,
it is possible
to modulate their physical and functional properties as required in optical and electronic
devices, biotechnology, and even in food packing~\cite{b2,b3,b4,b5,b6,b7,b8,b9}.
In this connection, extensive
studies of protein, peptide and other organic films have been performed
(see, for instance,
Refs.~\cite{b10,b11,b12,b13,b14,b15,b16,b17}).

It is well known that miniaturization is the main tendency in developing of modern
electronic devices, and organic electronics is not an exception to this rule. A number of organic
microdevices has already been created (see, e.g., Refs.~\cite{b5,b18,b19,b20}).
It was noticed that with
decreasing thickness of an organic film to below a micrometer the problem of its stability
acquires a serious meaning. There are several contributions to the free energy of a film which
is responsible for its stability~\cite{b21}. One of these contributions, which is gaining in importance
with decreasing film thickness, is caused by the zero-point and thermal fluctuations of the
electromagnetic field.

The fluctuation-induced force acting between two closely spaced material surfaces is
the long-explored subject. At separations below a few nanometers it is of quantum but
nonrelativistic character and is usually called the van der Waals force~\cite{b22}. At larger
separations the effects of relativistic retardation come into play and a frequently used name
is the Casimir force~\cite{b23}. The van der Waals and Casimir forces are also known under a generic
name dispersion forces~\cite{b23a}. They can be calculated by using the Lifshitz theory~\cite{b23,b24}.
For organic films, including the protein ones, the dispersion forces have long been investigated by
many anthors (see, e.g., Refs.~\cite{b25,b26,b27,b28}).

The Lifshitz theory can be also used to calculate the fluctuation-induced free energy of a
freestanding in vacuum or deposited on a material substrate thin film \cite{b21}. For this purpose,
one should exploit the Lifshitz formula for a three-layer system and put equal to unity the
dielectric permittivities of two or one outer layers, respectively. Investigations along these
lines have been performed recently for metallic and dielectric films, both freestanding and
deposited on substrates~\cite{b29,b30,b31,b32,b33,b34}.
It was found that the fluctuation-induced free energy of
a film can be both negative and positive, i.e., the effect of electromagnetic fluctuations
may be both favorable and unfavorable to its stability. For organic films, however, the
fluctuation-induced contribution to the free energy was not investigated so far.

In this paper, we use the Lifshitz theory to calculate the free energy of a freestanding
and deposited on a substrate peptide films. Both cases of dielectric (SiO$_2$) and metallic (Au)
substrates are considered. The dielectric permittivity of a typical peptide is modeled over
the wide range of imaginary frequencies by means of  simple analytic representation. The
free energy per unit area of peptide film is calculated at room temperature as a function
of film thickness for different volume fractions of water contained in a film. It is shown
that for a freestanding peptide film the fluctuation-induced free energy is negative. With
increasing volume fraction of water in a film, the magnitude of the free energy increases
by making the film more stable. For peptide films deposited on a SiO$_2$ substrate the free
energy is negative for thicker than $100~$nm films and
its magnitude reaches the maximum value for the film
of some definite thickness. For thinner than $100~$nm film
the free energy  may vanish and even become
positive with further decreased film thickness. It is shown that for protein films deposited
on metallic substrates the fluctuation-induced free energy takes the positive values.
The free
energy again increases with increasing  volume fraction of water, but this makes a film
less stable. In all three cases, for films of more than $2~\mu$m thickness, the fluctuation-induced
free energy reaches the classical limit. In doing so, simple analytic representations for the
fluctuation-induced free energy are obtained.

The paper is organized as follows. In Sec.~II, main formulas for the
fluctuation-induced free energy of both freestanding and deposited on a substrate
composite films are represented.
Section~III contains modeling of the dielectric permittivity and calculations
of the free energy for a freestanding peptide film. In Sec.~IV, the fluctuation-induced free
energies of peptide films deposited on dielectric and metallic substrates are found. Section~V
contains our conclusions and discussion.

\section{General formalism for freestanding and  deposited
{\protect{\\}} on a substrate peptide films}

We consider the peptide film of thickness $a$ described by the frequency-dependent
dielectric permittivity $\varepsilon^{(f)}(\omega)$. Peptide and protein films
are usually the composite layers.
They are made up not of entirely peptide or protein but contain some volume fraction of a
plasticizer to ensure the required physical and functional properties~\cite{b3}.
One of the plasticizers discussed in the literature is water~\cite{b35,b36} which is
contained in organic films in any
case. Because of this, below we consider the dielectric permittivity of peptide film, $\varepsilon^{(f)}$,
as a combination of the dielectric permittivities of peptide in itself, $\varepsilon^{(p)}(\omega)$,
and of water,
$\varepsilon^{(w)}(\omega)$.

It is assumed that the peptide film is deposited on a dielectric or metallic substrate
described by the dielectric permittivity $\varepsilon(\omega)$. The substrate is considered as
a semispace
when calculating the fluctuation-induced free energy of a film. For the validity of this
approach, the dielectric substrate should be thicker than approximately $2~\mu$m \cite{b37} and
metallic substrate thicker than 100~nm \cite{b23}. If a peptide film is freestanding in vacuum,
one should put $\varepsilon(\omega)=1$ in all subsequent formulas. Assuming that the film is in thermal
equilibrium with the environment at temperature $T$, the fluctuation-induced free energy of
this film is given by the Lifshitz-type formula~\cite{b23,b24,b32}
\begin{eqnarray}
&&
\mathcal{F}(a)=\frac{k_BT}{2\pi}\sum_{l=0}^{\infty}
\vphantom{\sum}^{'}\int_0^{\infty}kdk
\label{eq1}\\
&&~~~
\times \sum_{\alpha}\ln\left[1-r^{(f, v)}_{\alpha}(i{\xi}_l, k)r^{(f, s)}_{\alpha}(i{\xi}_l, k)e^{-2ak^{(f)}{(i{\xi}_l, k)}}\right].
\nonumber
\end{eqnarray}
\noindent
Here, $k_B$ is the Boltzmann constant, $r_{\alpha}^{(f, v)}$
and $r_{\alpha}^{(f, s)}$ are the reflection
coefficients on the
boundary surfaces between the peptide film and vacuum and substrate, respectively. These
reflection coefficients are defined for two independent polarizations
of the electromagnetic field,
transverse magnetic ($\alpha={\rm TM}$) and transverse electric
($\alpha={\rm TE}$). They are calculated at
the pure imaginary Matsubara frequencies $\xi_l=2\pi k_BTl/\hbar$, $l=0,\, 1,\, 2,\,\ldots$, and $k$ is the projection
of the wave vector on the plane of a film. The prime on the first summation sign in Eq.~(\ref{eq1})
means that the term with $l=0$ should be taken with the weight $1/2$ and
\begin{equation}
  k^{(f)}(i\xi_l,k)=\left[k^2+
  \varepsilon^{(f)}(i\xi_l)\frac{\xi_l^2}{c^2}\right]^{1/2}.
\label{eq2}
\end{equation}
\noindent
The explicit expressions for the reflection coefficients are the following:
\begin{eqnarray}
&&
r^{(f, v)}_{\rm TM}(i\xi_l,k)=
\frac{k^{(f)}(i\xi_l, k)-\varepsilon^{(f)}(i\xi_l)k^{(v)}(i\xi_l, k)}{k^{(f)}(i\xi_l, k)+\varepsilon^{(f)}(i\xi_l)k^{(v)}(i\xi_l, k)},
\nonumber\\
&&
r^{(f, v)}_{\rm TE}(i\xi_l,k)=
\frac{k^{(f)}(i\xi_l, k)-k^{(v)}(i\xi_l, k)}{k^{(f)}(i\xi_l, k)+k^{(v)}(i\xi_l, k)},
\label{eq3}
\end{eqnarray}
\noindent
and
\begin{eqnarray}
  &&
r^{(f, s)}_{\rm TM}(i\xi_l,k)=\frac{\varepsilon(i\xi_l)k^{(f)}(i\xi_l, k)-\varepsilon^{(f)}(i\xi_l)k^{(s)}(i\xi_l, k)}{\varepsilon(i\xi_l)k^{(f)}(i\xi_l, k)+\varepsilon^{(f)}(i\xi_l)k^{(s)}(i\xi_l, k)},
\nonumber\\
&&
r^{(f, s)}_{\rm TE}(i\xi_l,k)=
\frac{k^{(f)}(i\xi_l, k)-k^{(s)}(i\xi_l, k)}{k^{(f)}(i\xi_l, k)+k^{(s)}(i\xi_l, k)},
\label{eq4}
\end{eqnarray}
where
\begin{eqnarray}
  &&
k^{(v)}(i\xi_l, k)=\left(k^2+\frac{\xi^2_l}{c^2}\right)^{1/2},
\nonumber\\
  &&
k^{(s)}(i\xi_l, k)=\left[k^2+\varepsilon(i\xi_l)\frac{\xi^2_l}{c^2}\right]^{1/2}.
\label{eq5}
\end{eqnarray}

Equations~(\ref{eq2})--(\ref{eq5}) depend on the dielectric permittivities of substrate,
$\varepsilon(i\xi)$, and of peptide
film, $\varepsilon^{(f)}(i\xi)$. A peptide film is the mixture of peptide itself with the dielectric permittivity $\varepsilon^{(p)}(i\xi)$, which consists of molecules of irregular shape,
and of water with the
dielectric permittivity $\varepsilon^{(w)}(i\xi)$ which, in our case, plays the role of a plasticizer.
For the dielectric permittivities of water and peptide one can use the Clausius-Mossotti equation
\cite{38a}
\begin{equation}
\frac{\varepsilon^{(w,p)}(i\xi)-1}{\varepsilon^{(w,p)}(i\xi)+2}=
\frac{4\pi}{3}N^{(w,p)}\alpha^{(w,p)}(i\xi),
\label{eq5a}
\end{equation}
\noindent
where $N^{(w,p)}$ and  $\alpha^{(w,p)}$ are the numbers of molecules of water or peptide
per unit volume and their polarizabilities, respectively. If $\Phi$ is the volume fraction of water
 in peptide film, then, assuming no volume change on mixing of randomly distributed peptide
molecules with water, the
permittivity of peptide film, $\varepsilon^{(f)}(i\xi)$, is obtained from the following mixing formula \cite{b38}:
\begin{eqnarray}
  \frac{\varepsilon^{(f)}(i\xi)-1}{\varepsilon^{(f)}(i\xi)+2}&=&
\frac{4\pi}{3}[\Phi N^{(w)}\alpha^{(w)}(i\xi)+
(1-\Phi) N^{(p)}\alpha^{(p)}(i\xi)]
\nonumber \\
&=&
  \Phi\frac{\varepsilon^{(w)}(i\xi)-1}{\varepsilon^{(w)}(i\xi)+2}+
  (1-\Phi)\frac{\varepsilon^{(p)}(i\xi)-1}{\varepsilon^{(p)}(i\xi)+2}.
\label{eq6}
\end{eqnarray}

In the next two sections, Eqs.~(\ref{eq1})--(\ref{eq5}) and (\ref{eq6})
are used to calculate
the fluctuation-induced free energy of the freestanding and deposited on material
substrates peptide films and to investigate
its dependence on the film thickness and the fraction of water contained in the film.

\section{Dielectric permittivity and free energy of a {\protect{\\}}  freestanding peptide film}

We consider the freestanding peptide film of thickness $a$ containing the volume fraction
of water $\Phi$. Thus, we put $\varepsilon(i\xi_l)=1$ in Eqs.~(\ref{eq1})--(\ref{eq5}).
To find the dielectric permittivity of
a film, $\varepsilon^{(f)}(i\xi_l)$, one needs to know the dielectric permittivities of water, $\varepsilon^{(w)}(i\xi_l)$, and of
peptide, $\varepsilon^{(p)}(i\xi_l)$.

There are several representations for the permittivity of distilled water in the literature
\cite{b22, b38, b39, b40} which lead to approximately equal calculation results for the dispersion force.
 Below we use the representation of Ref.~\cite{b40}
\begin{equation}
  \varepsilon^{(w)}(i\xi_l)=1+\frac{B}{1+\tau\xi_l}+\sum_{j=1}^{11}\frac{C_j}{1+(\frac{\xi_l}{\omega_j})^2+
  \frac{g_j\xi_l}{\omega^2_j}},
\label{eq7}
\end{equation}
where $B=76.8$ and $1/\tau=1.08\times10^{11}$~rad/s are the parameters of the Debye term
descri\-bing orientation of permanent dipoles. The oscillator terms in Eq.~(\ref{eq7})
with $j=1,\,2,\,\ldots,\,5$
correspond to infrared frequencies, whereas the terms with $j=6,\,\ldots,\,11$ describe the
contribution of ultraviolet frequencies. The values of the
oscillator strengths $C_j$, oscillator frequencies $\omega_j$ and
relaxation parameters $g_j$ are
presented in Table I. As a result, $\varepsilon^{(w)}(0)=81.2$.

The dielectric properties of various proteins and peptides have been investigated by many
authors (see e.g., Refs.~\cite{b41, b42, b43, b44, b45, b46, b47, b48, b49}). The obtained results are, however, not sufficient for
calculation of the free energy using the Lifshitz theory which requires detailed information
on the dielectric permittivity over a wide frequency region from zero to far ultraviolet.
Here we present simple analytic form for the dielectric permittivity of a peptide film using
the numerical results of Refs.~\cite{b50, b51} obtained for different peptides in the microwave
and ultraviolet frequency regions, respectively.

For sufficiently thick films considered below the most important contribution to the
fluctuation-induced free energy is given by the region from zero to microwave imaginary
frequencies. Because of this, as the basic peptide sample for our calculation, we choose
the electrically neutral 18-residue zinc finger peptide. The molecules of this peptide
have size of a few nanometers and an irregular shape. As was shown in Ref.~\cite{b50} by
means of molecular-dynamics simulation, within the investigated frequency region up to
microwave frequencies its dielectric properties are well described by the frequency-dependent
complex dielectric permittivity.

The proposed representation for the dielectric permittivity of our peptide
sample along the imaginary
frequency axis is
\begin{equation}
  \varepsilon^{(p)}(i\xi)=1+\varepsilon_D(i\xi)+\varepsilon_{\rm IR}(i\xi)+\varepsilon_{\rm UV}(i\xi),
\label{eq8}
\end{equation}
where the second, third, and fourth terms on the right-hand side of Eq.~(\ref{eq8}) describe the
contributions of microwave, infrared, and ultraviolet frequencies, respectively. Using the
numerical results in Fig.~8 of Ref.~\cite{b50}, obtained for the imaginary part of dielectric
permittivity in the microwave region, and the Debye form for orientation polarization,
we find
\begin{equation}
  \varepsilon_D(i\xi)=\frac{C_D}{1+\tau_D\xi},
\label{eq9}
\end{equation}
where $C_D=9.47$ and $1/\tau=2.46\times10^8$ rad/s. According to the results of Ref.~\cite{b50}, $\varepsilon^{(p)}(0)=15$.

Unfortunately, information about the dielectric properties of zinc finger peptide in the
infrared and ultraviolet frequency regions is not available. Within the ultraviolet frequency region,
however, the imaginary part of the frequency-dependent dielectric permittivity of cyclic
tripeptide RGD-4C was computed on the basis of first principles of quantum mechanics
in Ref.~\cite{b51}. Taking into account that the ultraviolet region makes rather small
contribution to the free energy for considered film thicknesses and that a molecule
of RGD-4C is rather similar to that of zinc finger peptide in both shape and size, one may
expect that a replacement of $\varepsilon_{\rm UV}$ contribution to Eq.~(\ref{eq8})
with the one computed for RGD-4C will not lead to major errors.

The numerical results in Fig.~\ref{fg6} of Ref. \cite{b51} for the imaginary part of the
 permittivity of RGD-4C peptide in ultraviolet region lead to the following oscillator representation:
\begin{equation}
  \varepsilon_{\rm UV}(i\xi)=\sum_{i=1}^{3}\frac{C^{(i)}_{\rm UV}}{1+
  \left(\frac{\xi}{\omega^{(i)}_{\rm UV}}\right)^2+\frac{g^{(i)}\xi}{(\omega^{(i)}_{\rm UV})^2}}.
\label{eq10}
\end{equation}
Here, the oscillator strengths $C^{(i)}_{\rm UV}=0.022, 0.020$, and 0.191, the oscillator frequencies $\omega^{(i)}_{\rm UV}=
5.18, 6.10$, and 12.5~eV, and the relaxation parameters $g^{(i)}=0, 0$, and 14.0~eV for $i=1,\, 2$,
and 3, respectively.

The contribution of infrared frequencies is modeled in the Ninham-Parsegian
representation
\begin{equation}
  \varepsilon_{\rm IR}(i\xi)=\frac{C_{\rm IR}}{1+\frac{\xi^2}{\omega^2_{\rm IR}}},
\label{eq11}
\end{equation}
where $C_{\rm IR}$ is determined from already known parameters
\begin{equation}
  C_{\rm IR}=\varepsilon^{(p)}(0)-C_D-\sum_{i=1}^{3}C^{(i)}_{\rm UV}=4.3
\label{eq12}
\end{equation}
and  typical value of the oscillator frequency is $\omega_{\rm IR}=6.28\times10^{13}$~rad/s \cite{b52}.

\begin{figure}[b]
\vspace*{-4cm}
\centerline{\hspace*{2.5cm}
\includegraphics{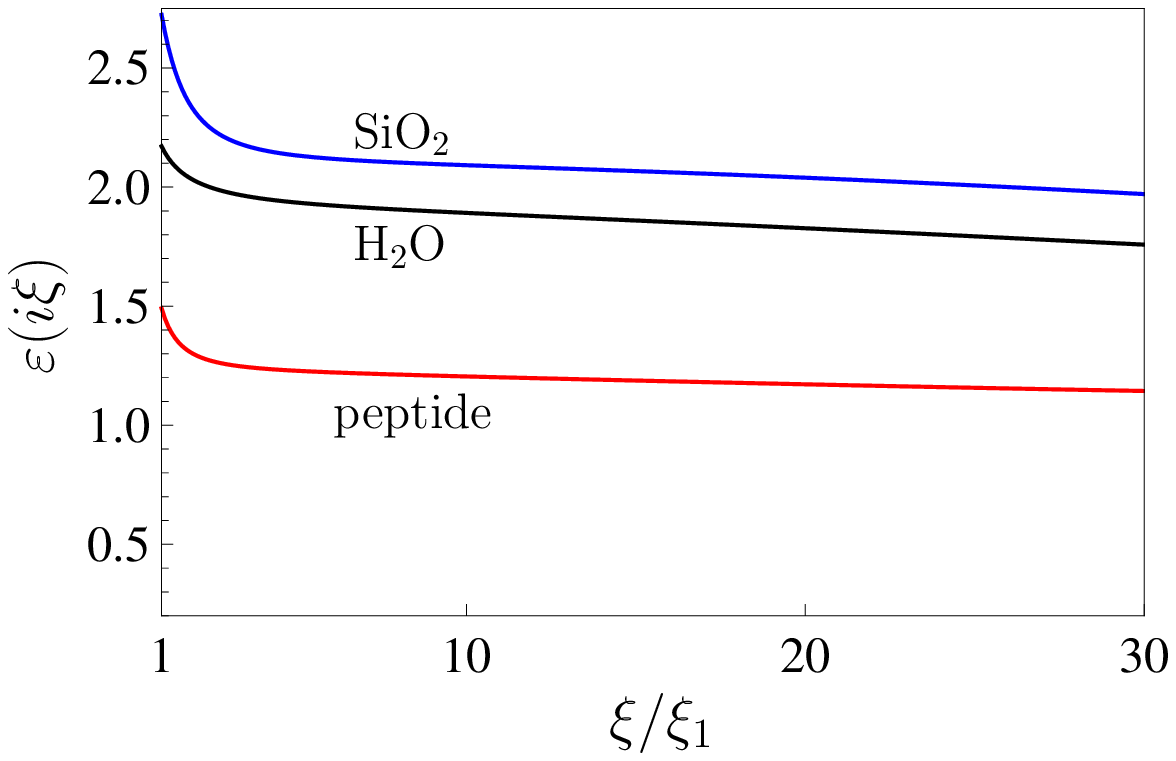}
}
\vspace*{-9.5cm}
\caption{\label{fg1}
The dielectric permittivities of peptide, water and silica glass are shown as the functions of imaginary frequency normalized to the first Matsubara frequency by the three lines plotted from bottom to top, respectively.
}
\end{figure}

In Fig.~\ref{fg1}, the dielectric permittivities of peptide and water given by Eqs.~(\ref{eq7}) and (\ref{eq8})--(\ref{eq11}),
respectively, are shown as the functions of imaginary frequency normalized to
the first Matsubara frequency over the interval from $\xi=\xi_1$
to $\xi=30\xi_1$. The top line in the same figure shows the dielectric permittivity of a silica glass
which is discussed in Sec.~IV as a substrate material. The values of all dielectric permittivities
at zero frequency are indicated in the text.

\begin{figure}[b]
\vspace*{-4cm}
\centerline{\hspace*{2.5cm}
\includegraphics{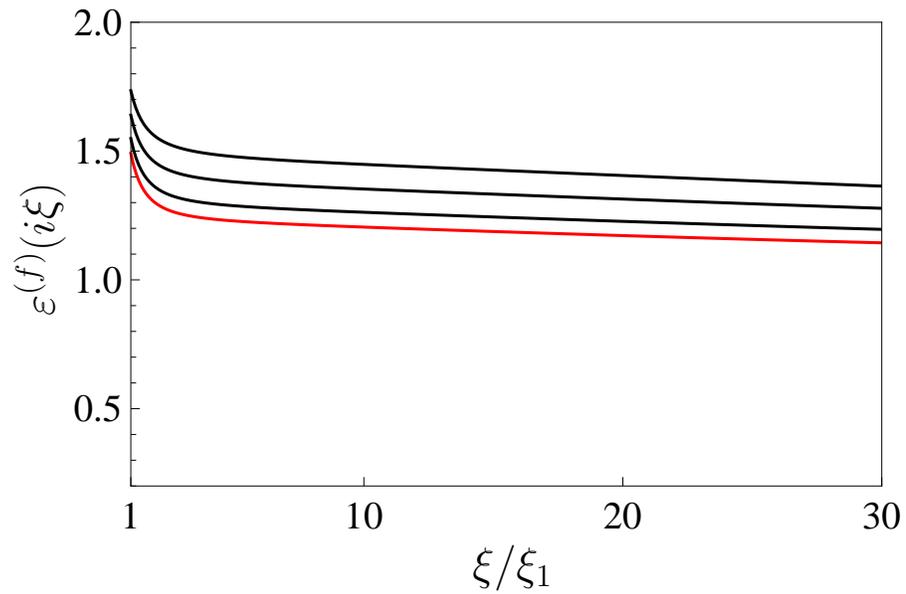}
}
\vspace*{-9.5cm}
\caption{\label{fg2}
The dielectric permittivities of peptide films containing 0, 0.1, 0.25, and 0.4 volume fractions of water are shown as the functions of imaginary frequency normalized to the first Matsubara frequency by the four lines plotted from bottom to top, respectively.
}
\end{figure}

The dielectric permittivities of peptide, $\varepsilon^{(p)}$, and of water, $\varepsilon^{(w)}$, have been combined by
using the mixing formula in Eq.~(\ref{eq6})
to obtain the dielectric permittivity of peptide film. In Fig.~\ref{fg2}
the permittivities of peptide film, $\varepsilon^{(f)}$, are shown as functions of the imaginary
frequency
normalized to the first Matsubara frequency by the four lines from bottom to top for the films
containing 0, 0.1, 0.25, and 0.4 volume fractions of water. The respective static permittivities
are equal to 15, 16.5, 19.2, and 22.9.

\begin{figure}[b]
\vspace*{-4cm}
\centerline{\hspace*{2.5cm}
\includegraphics{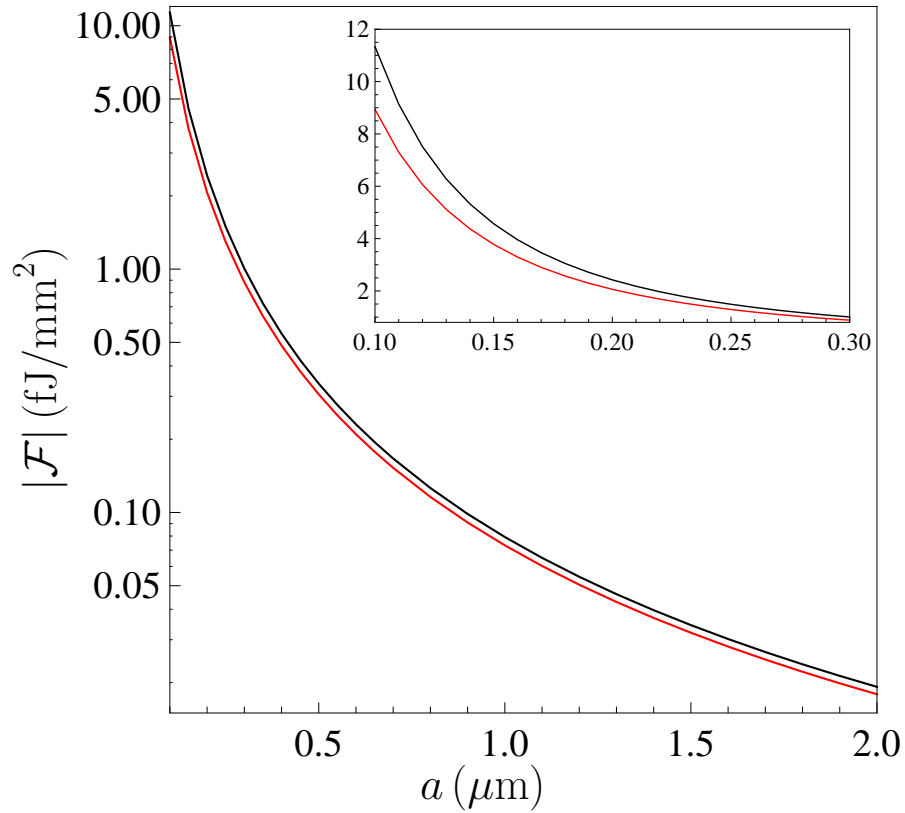}
}
\vspace*{-9.5cm}
\caption{\label{fg3}
The magnitudes of the fluctuation-induced free energies of peptide films containing 0 and 0.25 volume fractions of water are shown as the functions of film thickness by the two lines plotted from bottom to top, respectively. In an inset, the case of thinner films is illustrated using an uniform energy scale.
}
\end{figure}

Now we are in a position to calculate the fluctuation-induced free energy of a
freestanding peptide film. Computations were performed by Eqs.~(\ref{eq1})--(\ref{eq6})
at $T=300~$K with
$\varepsilon(i\xi_l)=1$ (i.e., with no substrate) for all-peptide film and for a film containing 0.25
volume fraction of water (i.e., with the permittivities $\varepsilon^{(f)}$ shown by the bottom and second
above the bottom lines in Fig.~\ref{fg2}). The computational results for the magnitudes of the free
energy as functions of film thickness are shown in Fig.~\ref{fg3} by the bottom and top lines for
all-peptide film  and for the film containing 0.25 fractions of water, respectively. In both cases
the free energy for films of any thickness is negative which is favorable to film stability.
In doing so, both the term with $l=0$, $\mathcal{F}^{(l=0)}$, in Eq.~(\ref{eq1}) and the sum of
all terms with
$l\geqslant1$, $\mathcal{F}^{(l\geqslant1)}$, are negative (compare with the case of peptide films
deposited on substrates
considered in Sec. IV).

The magnitude of fluctuation-induced free energy decreases monotonously with increasing film thickness. An increase of the volume fraction of water in the film results in
larger magnitude of the free energy for a film of the same thickness. As an example, for the
all-peptide film of $100~$nm thickness
$|\mathcal{F}|=8.938~\mbox{fJ/mm}^2$, but for the film of
the same thickness with $25\%$ of water one obtains $|\mathcal{F}|=11.33~\mbox{fJ/mm}^2$.
For the all-peptide and containing $25\%$ of
water films of 1 $\mu$m thickness we find $|\mathcal{F}|=0.07328$ and $0.07928~\mbox{fJ/mm}^2$, respectively.
For better visualization the case of thinner films from $100$ to $300~$nm thickness is shown
on an inset to Fig.~\ref{fg3} where the free energy is plotted in an uniform scale.

Note that the above computations have been performed with omitted conductivity of
peptide film at a constant current. The reason is that an inclusion of the dc conductivity of
dielectric materials in calculation of the dispersion interaction leads to contradictions with
the measurement data (see, e.g., Refs.~\cite{b23,b53,b54,b55}). In our case of peptide films, however,
this inclusion has only a slight effect on the obtained values of the free energy.

For sufficiently thick peptide film one can obtain rather simple analytic expression for the free
energy. In this case the dominant contribution to the free energy is given
by the term of Eq.~(\ref{eq1}) with $l=0$, whereas all terms with $l\geqslant1$ are exponentially small.
Preserving only the zero-frequency term in Eq.~(\ref{eq1}), one obtains
\begin{equation}
  \mathcal{F}(a)=\frac{k_BT}{4\pi}\int_{0}^{\infty}kdk\ln\left[1-r_{\rm TM}^{(f, v)^2}(0)e^{-2ak}\right],
\label{eq13}
\end{equation}
where, according to Eq.~(\ref{eq3}),
\begin{equation}
  r_{\rm TM}^{(f, v)}(0)=\frac{1-\varepsilon^{(f)}(0)}{1+\varepsilon^{(f)}(0)}.
\label{eq14}
\end{equation}
Integrating in Eq.~(\ref{eq13}), we find the free energy of peptide film in the so-called classical limit
\begin{equation}
  \mathcal{F}(a)=-\frac{k_BT}{16\pi a^2}{\rm Li}_{{3}}\left[{r_{\rm TM}^{(f, v)}}^2(0)\right],
\label{eq15}
\end{equation}
where ${\rm Li}_{{n}}(z)$ is the polylogarithm function.

For the freestanding all-peptide films the approximate expression~(\ref{eq15}) gives more than $99\%$
of the exact free energy for film thicknesses $a\geqslant1.5 \mu$m. In this case
$r_{\rm TM}^{(f, v)}(0)=-0.875$ and
\begin{equation}
  {\rm Li}_{{3}}\left[r_{\rm TM}^{(f, v)^2}(0)\right]=0.865.
\label{eq16}
\end{equation}
For the peptide films containing $25\%$ of water,
we have $r_{\rm TM}^{(f, v)}(0)=-0.901$ and
\begin{equation}
  {\rm Li}_{{3}}\left[r_{\rm TM}^{(f, v)^2}(0)\right]=0.927.
\label{eq17}
\end{equation}
In this case the approximate expression (\ref{eq15}) contributes more than $99\%$ of the exact free
energy for films with more or equal to $1.6~\mu$m thicknesses.

\section{Free energy of peptide films deposited on  dielectric
{\protect{\\}} and metallic substrates}

We begin with the case of peptide film deposited on dielectric substrate made of silica
glass SiO$_2$. The dielectric permittivity of SiO$_2$ along the imaginary frequency axis, $\varepsilon(i\xi)$,
was repeatedly used in calculations of the Casimir force~\cite{b23}. An analytic representation
for it in the Ninham-Parsegian representation is contained in Ref.~\cite{b38} (see the top line in
Fig.~\ref{fg1}).

\begin{figure}[b]
\vspace*{-4cm}
\centerline{\hspace*{2.5cm}
\includegraphics{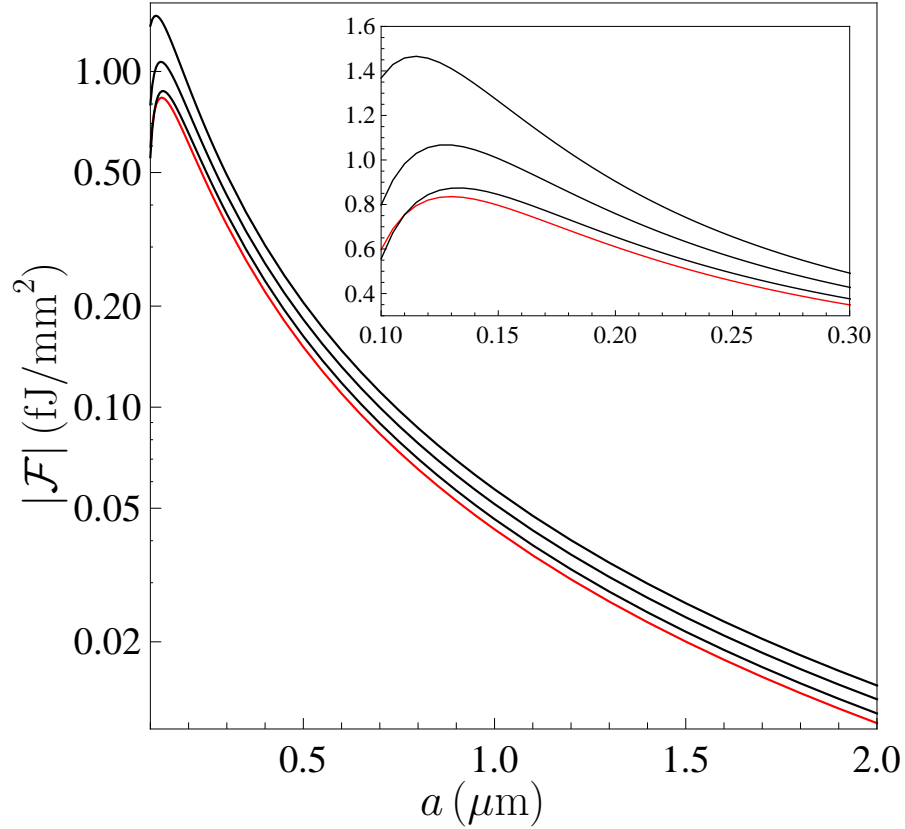}
}
\vspace*{-9.5cm}
\caption{\label{fg4}
The magnitudes of the fluctuation-induced free energies of peptide films, deposited on SiO$_2$
substrate, which contain 0, 0.1, 0.25, and 0.4 volume fractions of water, are shown as the
functions of film thickness by the four lines plotted from bottom to top, respectively.
In an inset, the case of thinner films is illustrated using an uniform energy scale.
}
\end{figure}

Computations of the fluctuation-induced free energy of peptide film deposited on SiO$_2$
substrate were done by Eqs.~(\ref{eq1})--(\ref{eq5}) at $T=300~$K.
The computational results for the
magnitudes of the free energy of all-peptide film and films containing 0.1, 0.25, and 0.4 volume
fractions of water as the functions of film thickness are shown in Fig.~\ref{fg4} by the four lines
plotted from bottom to top, respectively. In these computations the dielectric permittivities
of peptide films with different fractions of water shown by the four lines in Fig.~\ref{fg2} have been
used.

In the region of film thickness from $100~$nm to $2~\mu$m the free energy of peptide film
remains negative. However, unlike the case of a freestanding film, here we have $\mathcal{F}^{(l=0)}\textless0$
but $\mathcal{F}^{(l\geqslant 1)}>0$. Because of this, the free energy is a nonmonotonous function
of the film thickness. Intuitively, this behavior can be explained by the fact that the dielectric
permittivity of SiO$_2$ is larger than the dielectric permittivities of both water and peptide
at all nonzero Matsubara frequencies (see Fig. 1). The static dielectric permittivities of both
water and peptide are, however, larger than of  SiO$_2$. This relationship
between the dielectric permittivities determines a nonmonotonous behavior of the free energy.

From Fig.~\ref{fg4} it is seen that the free energy reaches the maximum values equal to
0.835, 0.8739, 1.0674, and $1.4657~\mbox{fJ/mm}^2$ for the films containing 0, 0.1, 0.25, and 0.4 volume
fractions of water and having the thicknesses of 130, 135, 130, and 115~nm, respectively. For
better visualization, the region of film thickness in the vicinities of maximum free energy
is shown on an inset where the uniform energy scale is used. For the all-peptide film of $85~$nm
thickness the fluctuation-induced free energy vanishes, $\mathcal{F}=0$, and for $a<85$~nm one has $\mathcal{F}>0$. Here we do not consider so thin films because this would demand a more exact expression for the dielectric permittivity of peptide at
high frequencies for obtaining
reliable computational results. The point is that in the ultraviolet frequency
region the dielectric permittivity of the zinc finger peptide under consideration was
approximated in Sec.~III by that of the RGD peptide. As a result, for sufficiently thin films, where
the contribution of ultraviolet frequencies becomes dominant, the computed values of the
free energy of a film might be burdened by rather big error. Because of this it is also
not reasonable to apply suggested expressions for determination of the Hamaker
constant of peptide film, which corresponds to the nonrelativistic limit of the Lifshitz
formula, i.e., to separation distances (film thicknesses) up to only a few nanometers
~\cite{b23}.

\begin{figure}[b]
\vspace*{-8cm}
\centerline{\hspace*{2.5cm}
\includegraphics{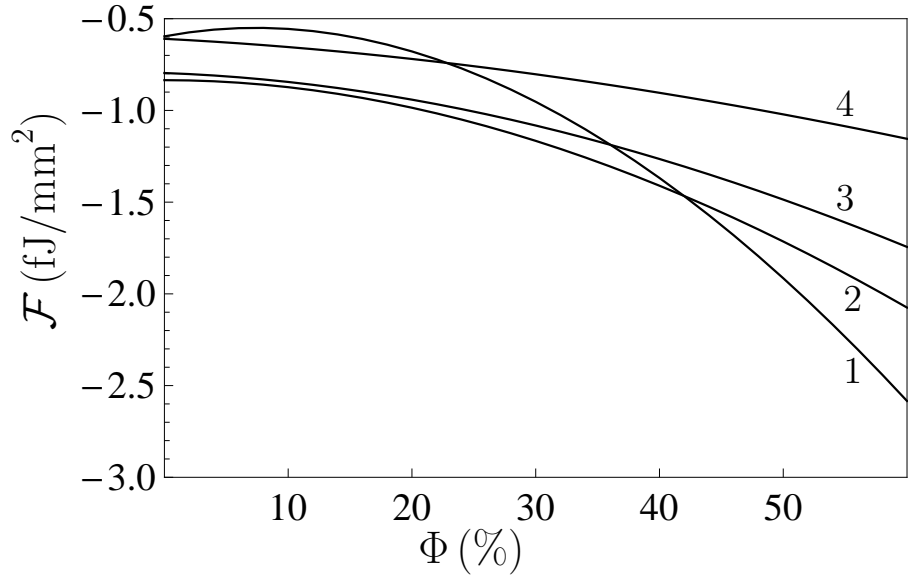}
}
\vspace*{-9cm}
\caption{\label{fg5}
The fluctuation-induced free energies of peptide films of thicknesses 100, 130, 150, and 200~nm
deposited on a SiO$_2$ substrate are shown as the functions of the fractions of water in the
film by the four lines labeled 1, 2, 3, and 4, respectively.
}
\end{figure}

The inset to Fig.~\ref{fg4} suggests that the dependence of the free energy of the fraction of
water $\Phi$ in a film deposited on a substrate may be more complicated than in the case of
a freestanding film. To confirm this guess, in Fig.~\ref{fg5} we present the computational results
for the free energy of peptide films with different thickness as the functions of $\Phi$. The lines
labeled 1, 2, 3, and 4 are plotted for the peptide films of 100, 130, 150, and 200~nm thickness
deposited on a SiO$_2$ substrate. As is seen in Fig.~\ref{fg5}, there is a peculiar interplay between
the film thickness and the fraction of water in their combined effect on the free energy of
peptide film deposited on a dielectric substrate.

As in the case of a freestanding film, for sufficiently thick peptide film deposited on a
dielectric substrate one can present simple analytic expression for the free energy. It is given
by the zero-frequency term of Eq.~(\ref{eq1})
\begin{equation}
  \mathcal{F}(a)=\frac{k_BT}{4\pi}\int_{0}^{\infty}kdk\ln\left[1-r_{\rm TM}^{(f, v)}(0)r_{\rm TM}^{(f, s)}(0)e^{-2ak}\right],
\label{eq18}
\end{equation}
where, according to Eq. (\ref{eq4}),
\begin{equation}
  r_{\rm TM}^{(f, s)}(0)=\frac{\varepsilon(0)-\varepsilon^{(f)}(0)}{\varepsilon(0)+\varepsilon^{(f)}(0)}
\label{eq19}
\end{equation}
and  $r_{\rm TM}^{(f, v)}(0)$ is given by Eq.~(\ref{eq14}).

Calculating the integral in Eq.~(\ref{eq18}), one obtains the free energy of peptide film in the classical
limit
\begin{equation}
  \mathcal{F}(a)=-\frac{k_BT}{16\pi a^2}
  {\rm Li}_{3}\left[r_{\rm TM}^{(f, v)}(0)r_{\rm TM}^{(f, s)}(0)\right].
\label{eq20}
\end{equation}
For deposited on a SiO$_2$ substrate peptide films with larger than 2.5~$\mu$m thickness,
Eq.~(\ref{eq20})
contributes more than $99\%$ of the free energy. For all-peptide films and for films containing
0.1, 0.25, and 0.4 volume fractions of water, one obtains
\begin{equation}
  {\rm Li}_3\left[r_{\rm TM}^{(f, v)}(0)r_{\rm TM}^{(f, s)}(0)\right] =0.562,\, 0.600,\, 0.660,
  ~\text{and 0.724},
\label{eq21}
\end{equation}
respectively. This allows simple calculation of the fluctuation-induced free energy for
sufficiently thick peptide films deposited on a SiO$_2$ substrate by using Eq.~(\ref{eq20}).

Now we consider peptide film deposited on a metallic substrate. As a substrate material,
we choose Au which is often used in measurements of dispersion forces~\cite{b23, b53}.
So, starting from
this point and below, $\varepsilon(i\xi)$ in Eqs.~(\ref{eq1})--(\ref{eq5}) means the dielectric
permittivity of Au which is
obtained from the measured optical data, extrapolated down to zero frequency using either
the plasma or the Drude model, with the help of the Kramers-Kronig relation ~\cite{b23, b53}. In
several experiments (see review in Refs. ~\cite{b23, b53, b56} and more modern measurements in
Refs. ~\cite{b57, b58,b59,b60}) it was shown that the use of the Drude model extrapolation taking into
account the relaxation properties of free electrons results in a dramatic contradiction with
the measurement data, whereas the plasma model extrapolation brings the theory in perfect
agreement with the data. Below we use the dielectric permittivity of Au obtained by means
of the plasma model extrapolation. Note, however, that here the Drude-plasma choice leads
to only minor differences between the obtained free energies of peptide film because the
metallic layer serves as only a substrate.

\begin{figure}[b]
\vspace*{-4cm}
\centerline{\hspace*{2.5cm}
\includegraphics{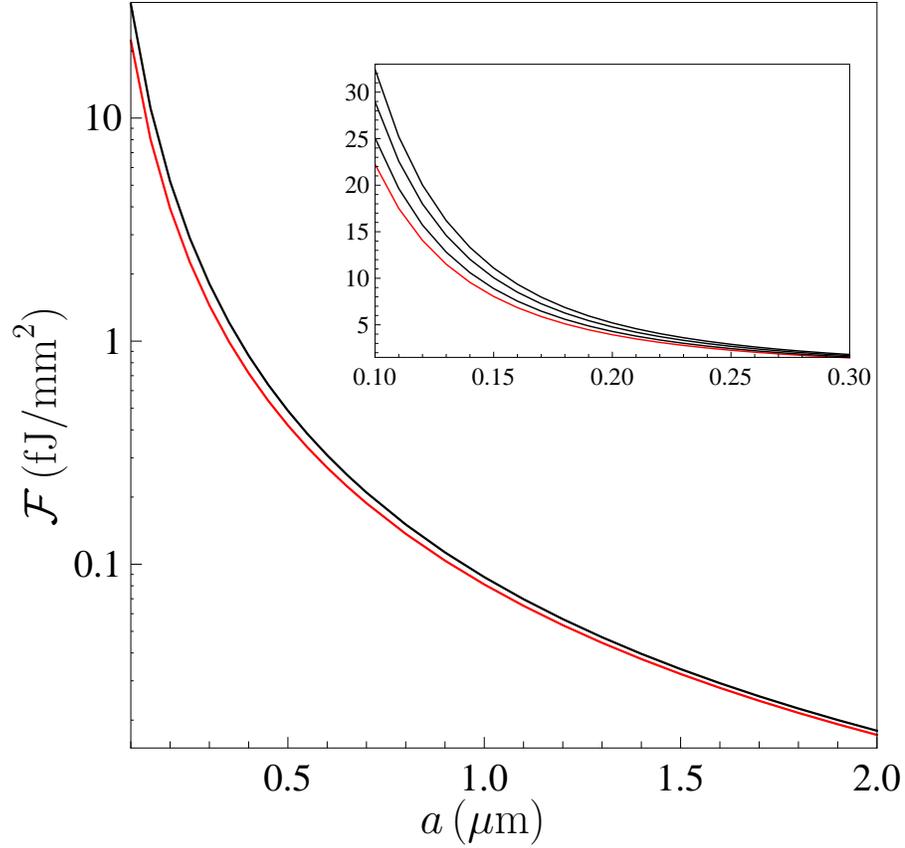}
}
\vspace*{-9.5cm}
\caption{\label{fg6}
The fluctuation-induced free energies of peptide films, deposited on Au substrate, which contain
0 and 0.4 volume fractions of water, are shown as the functions of film thickness by the bottom
and top lines, respectively. In an inset, the free energy of peptide films containing 0, 0.1,
0.25, and 0.4 volume fractions of water are shown by the four lines plotted from bottom to top,
respectively, using an uniform energy scale.
}
\end{figure}

Computations of the fluctuation-induced free energy have been performed by
Eqs.~(\ref{eq1})--(\ref{eq5})
at $T=300~$K for all-peptide film and for films containing 0.1, 0.25, and 0.4 volume
fractions of water deposited on Au substrate. The computational results for all-peptide
film and for a film containing 0.4 fraction of water as the functions of film thickness are
shown in Fig.~\ref{fg6} by the bottom and top lines, respectively.
As is seen in Fig.~\ref{fg6}, in both
cases the free energy of a film is positive. The radical difference from the case of a dielectric
substrate is that here both contributions to the free energy, $\mathcal{F}^{(l=0)}$ and $\mathcal{F}^{(l\geqslant1)}$, are positive.
Because of this, the fluctuation-induced contribution to the free energy of peptide films
deposited on metallic substrates makes them less stable.

The range of smaller film thicknesses is shown on the inset
to Fig.~\ref{fg6} where the uniform energy scale
is used. Here, the computational results for the free energy of all-peptide film and for
films with 0.1, 0.25, and 0.4 fractions of water are shown as the functions of film thickness
by the four lines plotted from bottom to top, respectively. From Fig.~\ref{fg6} it is seen that in
the case of metallic substrate the free energy of peptide films increases monotonously with
increasing volume fraction of water in the film. By way of example, for the all-peptide film
of $100~$nm thickness deposited on Au substrate we have $\mathcal{F}=22.25~\mbox{fJ/mm}^2$, but for a
similar film containing 25$\%$ of water  $\mathcal{F}=28.97~\mbox{fJ/mm}^2$. For 1~$\mu$m thick all-peptide and
containing 25$\%$ of water films one obtains $\mathcal{F}=0.08116$ and $0.08539~\mbox{fJ/mm}^2$, respectively.
This is qualitatively similar to the case of a freestanding peptide film, but in the presence
of metallic substrate the magnitudes of the fluctuation-induced free energy are larger.

\begin{figure}[b]
\vspace*{-8cm}
\centerline{\hspace*{2.5cm}
\includegraphics{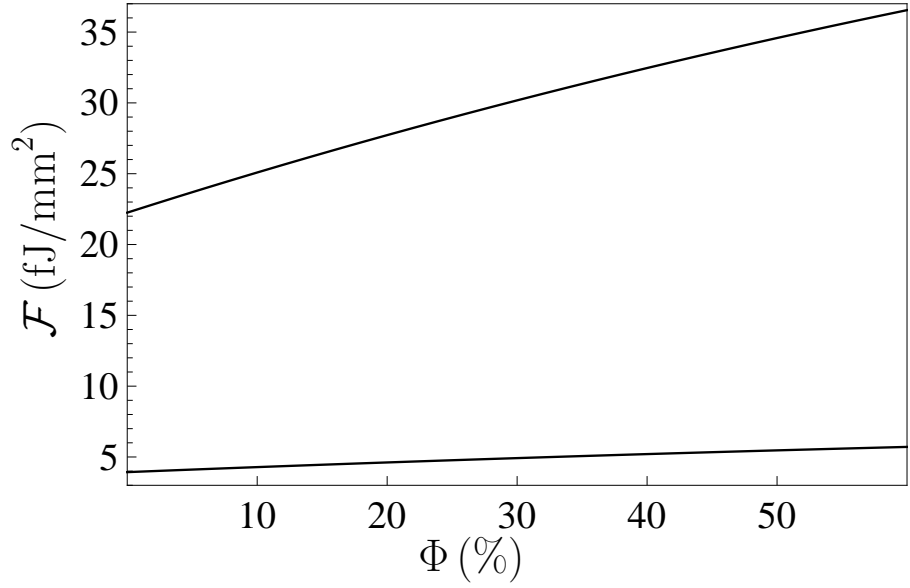}
}
\vspace*{-9cm}
\caption{\label{fg7}
The fluctuation-induced free energies of peptide films of thicknesses 200 and 100 nm deposited on Au substrate
are shown as the functions of the fraction of water in the film by the bottom and top lines, respectively.
}
\end{figure}

We have also computed the free energy of a peptide film deposited on Au substrate as a
function of the volume fraction of water $\Phi$ contained in the film. The computational results
are shown in Fig.~\ref{fg7} as the functions of $\Phi$ by the bottom and top lines plotted for the films
of 200 and $100~$nm thickness. Here, as opposed to Fig.~\ref{fg5} plotted for a dielectric substrate,
the free energy increases monotonously with $\Phi$.

In the end of this section, we consider the case of sufficiently thick peptide films when
the major contribution to the free energy~(\ref{eq1}) is given by the term with $l=0$. In this case
the free energy is again presented by Eq.~(\ref{eq18}) where $r_{\rm TM}^{(f, v)}(0)$ is expressed by Eq.~(\ref{eq14}) and
$r_{\rm TM}^{(f, s)}(0)=1$ in accordance to Eq.~(\ref{eq19}) because $\varepsilon(0)=\infty$.
The integration in Eq. (\ref{eq18})
results in
\begin{equation}
  \mathcal{F}(a)=-\frac{k_BT}{16\pi a^2}{\rm Li}_{3}\left[r_{\rm TM}^{(f, v)}(0)\right].
\label{eq22}
\end{equation}
For peptide films of more than 2.5 $\mu$m thickness deposited on Au substrate Eq.~(\ref{eq22}) gives
approximately 99$\%$ of the fluctuation-induced free energy. Thus, for sufficiently thick all-
peptide and containing 0.1, 0.25, and 0.4 volume fractions of water films deposited on Au
substrate we have
\begin{equation}
  {\rm Li}_{3}\left[r_{\rm TM}^{(f, v)}(0)\right]=-0.798,\,
  -0.806,\, -0.820,~\text{and}~-0.832,
\label{eq23}
\end{equation}
respectively, and one can calculate the fluctuation-induced free energy using Eq.~(\ref{eq22}).

\section{Conclusions an discussions}
In this paper, we have considered the contribution to the free energy of peptide films
which is induced by the zero-point and thermal fluctuations of the electromagnetic field.
This contribution may become relatively large for sufficiently thin films and should be taken
into account in the balance of energies responsible for the stability of a film. Taking into
account that thin peptide, protein and other organic films are already used as constituent
parts of various microdevices discussed in Sec. I, the role of quantum fluctuations in such
films deserves attention.

The formalism allowing calculation of the fluctuation-induced free energy of peptide films,
both freestanding and deposited on a substrate, is based on the Lifshitz theory of dispersion
forces. Application of this formalism requires knowledge of the dielectric permittivities of
all involved materials over the wide ranges of imaginary frequencies. Using the available
numerical data for the imaginary parts of dielectric permittivities of peptides, we have
devised an analytic expression for the dielectric permittivity of a typical peptide along the
imaginary frequency axis. This permittivity was combined with the dielectric permittivity
of different volume fractions of water to obtain the dielectric permittivity of a peptide film.

The numerical computations of the fluctuation-induced free energy have been performed
at room temperature for a freestanding peptide film and for films deposited on
dielectric (SiO$_2$) and
metallic (Au) substrates. It is shown that the free energy of a freestanding film is always
negative and, thus, contributes to the film stability. The magnitude of the free energy
decreases with increasing film thickness, but increases with increasing fraction of water
contained in
the film.

For peptide films deposited on a SiO$_2$ substrate, the fluctuation-induced free energy
is shown to be a nonmonotonous function of the film thickness. With increasing thickness
of a film, the magnitude of the free energy reaches its maximum value (for thicknesses in
the region from 115 to $135~$nm depending on the fraction of water) and then decreases. For
thinner than $100~$nm films deposited on a SiO$_2$ substrate the fluctuation-induced free energy
may vanish  (for $85~$nm thick all-peptide film) and even become positive. The dependence of
the free energy on the fraction of water in the film deposited on a SiO$_2$ substrate demonstrates
a nontrivial character dictated by the film thickness. Intuitively, this can be explained by the fact
that with increasing fraction of water the dielectric permittivity of a film increases at all
Matsubara frequencies. At zero Matsubara frequency this increase is, however, relatively
larger than at nonzero Matsubara frequencies. For different film thicknesses the relative
contribution of zero and nonzero Marsubara frequencies to the free energy varies resulting
in a nontrivial dependence on the fraction of water.

The case of peptide films deposited on metallic (Au) substrate possesses important special
features. According to our results, the fluctuation-induced free energy of peptide films in
this case is always positive and, thus, makes the film less stable. The free energy decreases
monotonously with increasing film thickness and increases with increasing fraction of water
in the film.

To understand the role of fluctuation-induced phenomena in stability of peptide films,
it would be interesting to compare the computed free energy with cohesive energies that
are responsible for stability of these films. Taking into account a lack of information for
specific films under consideration, it is possible to make only a qualitative estimation.
Thus, for the mussel-inspired peptide films of 3--5~nm thickness deposited on substrates
made of different materials the cohesive energy was measured to be of order
$1~\mbox{mJ/m}^2$ \cite{b61,b62,b63,b64}. This value is contributed by the ionic
or electrostatic interactions, hydrogen bonding, hydrophilic forces, fluctuation
phenomena etc. To make a comparison, we scale, e.g., the fluctuation-induced free energy
of our peptide film of 100~nm thickness with 10\% fraction of water deposited on Au
substrate (${\cal F}=25.08~\mbox{nJ/m}^2$) to 3--5~nm thickness using the scaling
law $\sim a^{2.56}$ and obtain ${\cal F}$ varying from 0.2 to $0.05~\mbox{mJ/m}^2$,
respectively. Thus, the fluctuation-induced free energy may contribute from 5 to 20\% of the
cohesive energy of a film.

For sufficiently thick peptide films, simple analytic expressions for their
fluctuation-induced free energy are obtained. These expressions give 99$\%$ of the free energy
for thicker
than 1.5--$1.6~\mu$m freestanding films and for thicker than $2.5~\mu$m films deposited on
dielectric or metallic substrates.

In the future, it would be interesting to assess an applicability of the obtained results to
different kinds of peptide and, even wider, protein and organic films of various constitutions.
This might be helpful for resolving the problem of film stability when developing the next
generation of organic microdevices with further shrinked dimensions.

\section*{Acknowledgments}

The work of V.~M.~M.~was partially supported by the Russian Government Program
of
Competitive Growth of Kazan Federal University.

\newpage
\begingroup
\squeezetable
\begin{table}
\caption{The values of the oscillator strengths $C_j$, oscillator frequencies
$\omega_j$ and relaxation parameters $g_j$ for the dielectric permittivity of
distilled water.}
\begin{ruledtabular}
\begin{tabular}{rlll}
$j$&$~C_j$ & $\omega_j~$(rad/s) &$g_j~$(rad/s) \\
\hline
1 & 1.46 & $0.314\times 10^{14}$ & $2.29\times 10^{13}$  \\
2 & 0.737 & $1.05\times 10^{14}$ & $5.78\times 10^{13}$  \\
3 & 0.152 & $1.40\times 10^{14}$ & $4.22\times 10^{13}$  \\
4 & 0.0136 & $3.06\times 10^{14}$ & $3.81\times 10^{13}$  \\
5 & 0.0751 & $6.46\times 10^{14}$ & $8.54\times 10^{13}$  \\
6 & 0.0484 & $1.25\times 10^{16}$ & $0.957\times 10^{15}$ \\
7 & 0.0387 & $1.52\times 10^{16}$ & $1.28\times 10^{15}$ \\
8 & 0.0923 & $1.73\times 10^{16}$ & $3.11\times 10^{15}$ \\
9 & 0.344 & $2.07\times 10^{16}$ & $5.92\times 10^{15}$ \\
10 & 0.360 & $2.70\times 10^{16}$ & $11.1\times 10^{15}$ \\
11 & 0.0383 & $3.83\times 10^{16}$ & $8.11\times 10^{15}$
\end{tabular}
\end{ruledtabular}
\end{table}
\endgroup
\end{document}